\documentclass[prl,aps,reprint,twocolumn,footinbib,showpacs,superscriptaddress]{revtex4-1}
\usepackage{graphicx}
\usepackage{amssymb}
\usepackage{amsmath}
\usepackage{times}

\begin{document}

\title{From quasiperiodic partial synchronization to collective chaos 
in populations of inhibitory neurons with delay}
\author{Diego Paz\'o}
\affiliation{Instituto de F\'{\i}sica de Cantabria (IFCA), CSIC-Universidad de 
Cantabria, 39005 Santander, Spain}
\author{Ernest Montbri\'o}
\affiliation{Center for Brain and Cognition,
Department of Information and Communication Technologies,
Universitat Pompeu Fabra, 08018 Barcelona, Spain}
\date{\today}

\begin{abstract}
Collective chaos is shown to emerge, via a period-doubling cascade, from 
quasiperiodic partial synchronization in a population of
identical inhibitory neurons with delayed global coupling.
This system is thoroughly investigated by means of 
an exact model of the macroscopic dynamics, 
valid in the thermodynamic limit.
The collective chaotic state is reproduced numerically 
with a finite population, and 
persists in the 
presence of weak heterogeneities.      
Finally, the relationship of the model's dynamics with fast neuronal oscillations 
is discussed.
\end{abstract}

\maketitle

Electrical measurements of brain activity display a broad spectrum of 
oscillations, reflecting the complex coordination 
of spike discharges across large neuronal 
populations~\cite{Wan10}. 
A particularly fruitful theoretical framework for investigating neuronal rhythms  
is to model networks of neurons as populations of heterogeneous oscillators~
\cite{Win80,HI97,ACN16}. 
These models exhibit a prevalent transition from 
incoherence to partial coherence, when 
a fraction of the oscillators becomes entrained to a common frequency. As a result
a macroscopic oscillatory mode appears with the same 
frequency as that of the synchronized cluster~\cite{Win80,Kur84}. 

Yet, even populations of globally coupled \textit{identical}
oscillators are capable of exhibiting a much wider diversity 
of complex oscillatory states, see \cite{PikRos15} for a recent survey. 
In general, this is due to 
the complexity of the coupling functions and of the individual 
oscillators. A relevant example is the so-called \textit{quasiperiodic partial 
synchronization} (QPS), which has been extensively investigated in  
networks of excitatory leaky integrate-and-fire (LIF) 
neurons~\cite{Vre96,MP06,OPT10,LOP+12,BSV+14},
as well as in populations of limit-cycle oscillators and 
phase oscillators~\cite{VD03,RP07,PR09,temirbayev12,temirbayev13,RP15,PR15}. 
In QPS, the network sets into a nontrivial dynamical regime 
in which oscillators display quasiperiodic dynamics 
while the collective observables oscillate periodically. 
Remarkably, the period of these oscillations differs 
from the mean period of the individual oscillators. 
As pointed out recently~\cite{RP15}, this interesting property of QPS 
is shared by the \emph{collective chaos} observed
in populations of globally coupled limit-cycle oscillators
~\cite{HR92,NK93,NK94,NK95,TGC09,TCG+11,KGO15}. 
Here, the collective chaotic mode is typically  
accompanied by microscopic chaotic dynamics at the level of the individual oscillators. 
However, as noticed in~\cite{NK93},
populations of limit-cycle oscillators  
may also display
pure collective chaos without trace of orbital instability at the microscopic level.
In this state the coordinates of the oscillators fall on a smooth closed curve and 
no mixing occurs, what points to the existence of collective chaos in 
populations of oscillators governed by a single phase-like variable.

In this Letter we uncover the spontaneous emergence of pure collective chaos from
QPS, via a cascade of period-doubling bifurcations. 
Notably, this is found  
in a simple population of identical integrate-and-fire oscillators with
time-delayed pulse coupling, which 
is thoroughly analyzed within the framework of the so-called Ott-Antonsen 
theory~\cite{OA08,LBS13,PM14,MPR15}.
Moreover, we show that pure collective chaos persists when 
weak heterogeneities are considered. This suggests that 
certain forms of irregular collective motion observed in large networks of 
heterogeneous LIF neurons with delay~\cite{LP10} may be 
already found for identical neurons. 

We investigate a model consisting of a population of  
$N\gg1$ neurons, with membrane potentials $\{V_j\}_{j=1,\ldots,N}$.
The evolution of $V_j$ is governed by the   
so-called quadratic integrate-and-fire (QIF) 
model, which obeys the nonlinear differential equation
~\cite{EK86,LRN+00,Izh07}
\begin{equation}
\tau {\dot V}_j= V_j^2+ I_j, 
\label{qif} 
\end{equation}
where $\tau$ is the neuron's membrane time constant.
When $V_j$ reaches the value $V_p$, 
the QIF neuron emits a spike, 
and $V_j$ is reset to $V_r$. 
Thereafter we consider $V_p=-V_r=\infty$~
\footnote{See Supplementary Material [url] for details of the numerical implementation.}. 
In this case the model \eqref{qif} can be exactly transformed to a 
phase model called theta-neuron~\cite{EK86,Izh07}.
The external inputs  $I_j$ have the form 
\begin{equation}
I_j=\eta_j + J \, s_D,
\label{etaj}
\end{equation}
where parameters $\eta_j$ determine the 
dynamics of each uncoupled neuron, $J=0$:
Those neurons with $\eta_j<0$ are excitable, 
whereas neurons with $\eta_j>0$ behave as self-sustained oscillators with
period, or interspike interval $\mathrm{ISI}_j=\pi \tau/\sqrt{\eta_j}$.
In Eq.~\eqref{etaj}, the delayed mean activity $s_D\equiv s(t-D)$ 
is defined summing the spikes of all neurons:
\begin{eqnarray}
s_D=  \frac{\tau}{N \tau_s} \sum_{j=1}^N  \sum_{k} \int_{t-D-\tau_s}^{t-D} \delta (t'-t_j^{k}) \, dt'.
\label{s}
\end{eqnarray}  
In this equation, $t_j^k$ is the time of the $k$th spike of $j$th neuron, and
$\delta (t)$ is the Dirac delta function.
We assume the thermodynamic limit $N\to\infty$, so that
a second limit in the temporal window $\tau_s\to0$ leads to the 
relationship $s_D=\tau r_D$,
where $r_D\equiv r(t-D)$ is the time-delayed firing rate, i.e.~the population-averaged number of spikes per unit time.
The strength of the interactions is controlled in Eq.~\eqref{etaj} 
by the synaptic weight constant $J$, which can be either positive or 
negative for excitatory or inhibitory synapses, respectively.

%%%%%%%%%%%%%%%%%%
\begin{figure}
\centerline{\includegraphics[width=70mm,clip=true]{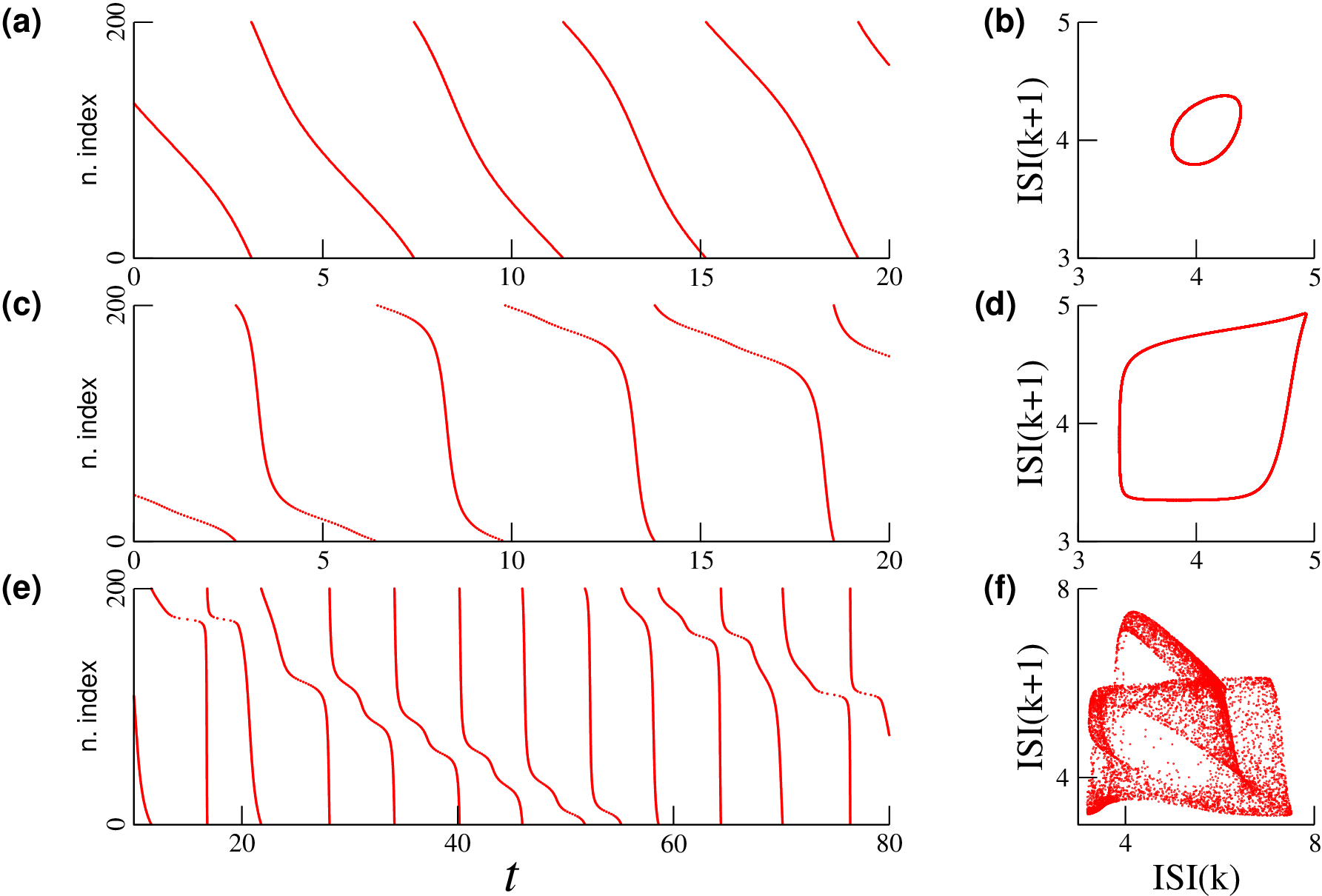}}
\caption{Quasiperiodic partial synchronization (a-d) 
and collective chaos (e,f) 
in an inhibitory network of $N=1000$ identical QIF neurons with 
$\eta_j=\bar\eta=\tau=1$, $\tau_s=10^{-3}$, and: 
(a,b) $D=2.5$, $J=-1.65$, 
(c,d) $D=2.5$, $J=-1.85$, and 
(e,f) $D=3$, $J=-3.8$.
(a,c,e) Raster plots of 200 randomly selected neurons. 
Dots corresponds to a firing events, and neurons
are indexed according to their firing order.
(b,d,f) Return plots for $10^4$ interspike intervals
$\mathrm{ISI}_j(k)=t_j^{k+1}-t_j^k$ of an arbitrary neuron $j$.} 
\label{Fig1}
\end{figure}
%%%%%%%%%%%%%%%%%

We start performing numerical simulations of an inhibitory 
($J<0$) population of identical neurons with $\eta_j=\bar \eta>0$. 
In panels (a) and (c) of Fig.~\ref{Fig1},
showing raster plots for two values of $J$, the system exhibits QPS.
In fact, the return plots in panels (b,d) show a closed line
indicating quasiperiodic single-neuron dynamics, see \cite{Vre96}. 
Remarkably, for certain values of the time delay $D$, see  Fig.~\ref{Fig1}(e,f), 
increasing inhibition leads to a different macroscopic state, 
where neurons exhibit irregular dynamics whereas the 
macroscopic dynamics is chaotic, as shown below. 
%(see below).

Where and how QPS and collective chaos emerge is investigated next.
To this aim, we follow~\cite{MPR15} and using the Ott-Antonsen theory
(by means of a Lorentzian ansatz) derive the so-called 
firing-rate equations (FREs) governing the dynamics of the firing rate $r$,
and the population's mean membrane potential $v$.
Considering that currents $\eta_j$ are distributed according to 
a Lorentzian distribution of  half-width $\Delta$, centered at $\bar \eta$,
$g(\eta)=(\Delta/\pi)[(\eta-\bar\eta)^2+\Delta^2]^{-1}$,
we obtain a system of one ordinary and one delay 
differential equations~\footnote{See 
Supplementary Material [url] for the exact derivation of the FREs corresponding to Eqs.~\eqref{qif}-\eqref{etaj}.}  
\begin{subequations}
\label{fre}
\begin{eqnarray}
\tau \dot r &=& \frac{\Delta}{\pi \tau} + 2  r v, \\ 
\tau \dot v &=&   v^2 +   \bar \eta + J \tau r_D -\tau^2 \pi^2 r^2  , \label{freb}
\end{eqnarray}
\end{subequations}
which exactly describe the macroscopic dynamics of the system 
in the infinite $N$ limit 
\footnote{For identical neurons the dynamics of the model
is degenerate (and described by the Watanabe-Strogatz theory \cite{WS94}), 
but the presence of a tiny amount of noise attracts the dynamics to the Lorentzian
manifold, making Eq.~\eqref{fre} with $\Delta=0$ asymptotically correct \cite{vlasov}.
Accordingly, the initial conditions for the numerical simulations in Figs.~1 and 2 
are taken to represent a Lorentzian density with arbitrary values of the center
$v$ with half-width $\pi r$: $V_j(0)= v +\pi r \tan[(\pi/2)(2j-N-1)/(N+1)]$.}.
Hereafter we set $\tau=\bar\eta=1$ in Eq.~\eqref{fre}, 
without lack of generality
\footnote{This can always be achieved (for $\bar\eta>0$) under
rescaling of time $\tilde t=t \sqrt{\bar\eta}/\tau$, the
variables $\tilde r=r \tau /\sqrt{\bar \eta}, 
\tilde v = v /\sqrt{\bar \eta}$ and the parameters
$\tilde J=J/\sqrt{\bar \eta}$, $\tilde D=D  \sqrt{\bar \eta}/ \tau$,
$\tilde\Delta= \Delta/\bar\eta$.}.

%%%%%%%%%%%%%%%%%%
\begin{figure}
\centerline{\includegraphics[width=70mm,clip=true]{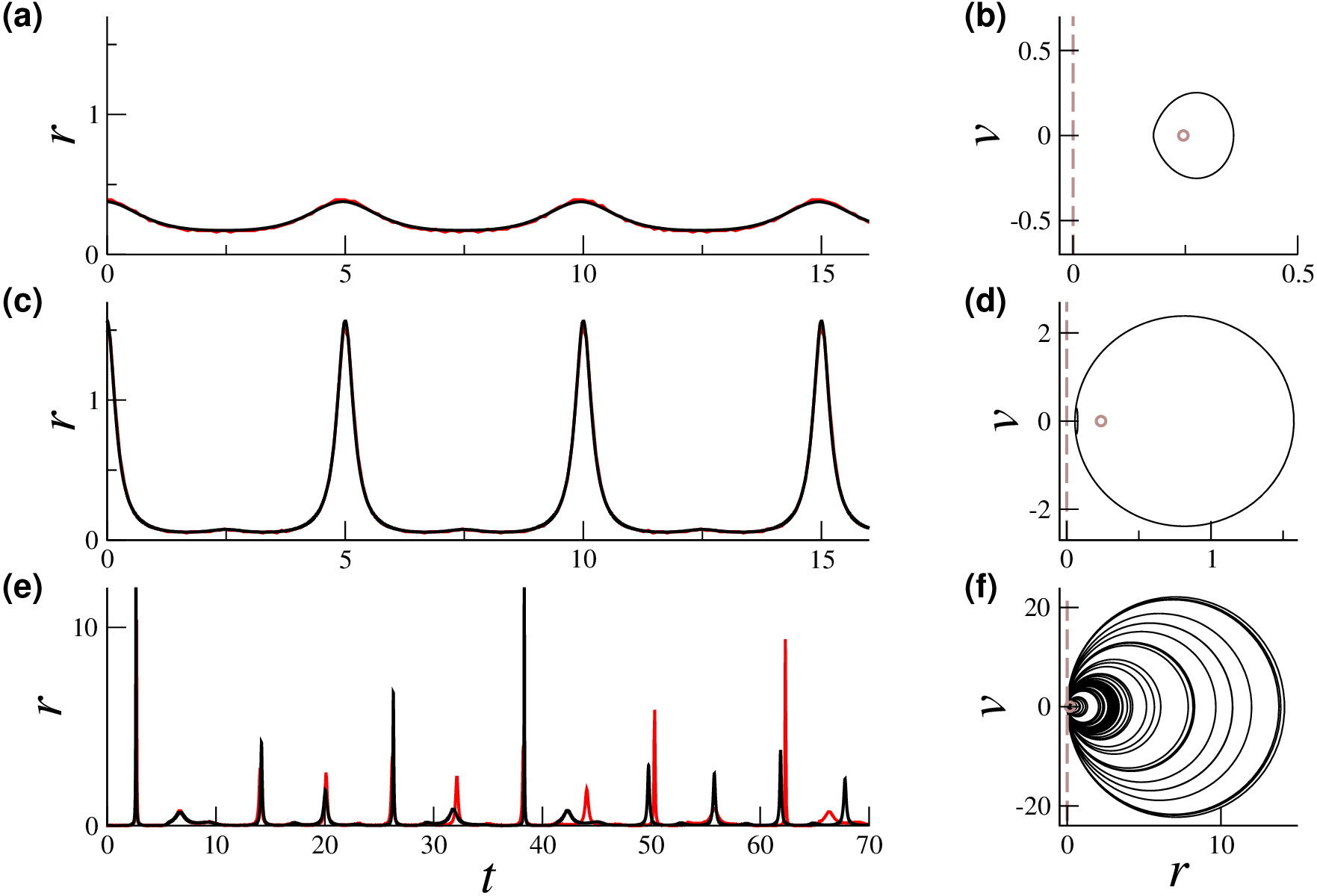}}
\caption{Macroscopic dynamics of  
quasiperiodic partial synchronization (a-d), and  collective chaos (e,f); same parameters as in Fig.~\ref{Fig1} are used. 
Black curves are obtained integrating the FREs~\eqref{fre}, with $\Delta=0$.
Red curves in (a,c,e) are the firing rates obtained from numerical simulation
of the population, Eqs.~\eqref{qif}-\eqref{s}. 
Right panels (b,d,f): Phase portraits of the FREs. 
The unstable fixed point (brown circle)
and the unstable orbit (brown, dashed line) correspond to incoherence and full synchronization, respectively.
The three largest Lyapunov exponents of the chaotic attractor in panel (f) are $\{0.055,
0,-0.232\}$.}
\label{Fig2}
\end{figure}
%%%%%%%%%%%%%%%%%

Figures~\ref{Fig2}(a,c,e) display the time series of the population
firing rate, using the parameters of Fig.~\ref{Fig1},
for both the network of spiking neurons \eqref{qif} and the FREs \eqref{fre}.
The attractor of the FREs in Fig.~\ref{Fig2}(b,d,f) is in perfect agreement
with the global behavior of the population.
Figures~\ref{Fig1}(b,d) and \ref{Fig2}(a,c) display
the fingerprint of QPS: 
oscillations of the mean field, 
%(here, the firing rate $r$) 
with a different period (in the present case, longer) 
than the individual neurons ISIs. 
Noteworthy, the two oscillations shown in Figs.~\ref{Fig2}(a,c) 
have \emph{exactly} the same period: $T_1=2D$. 
This is the consequence of the symmetry of  
the limit cycle under $v\to-v$, see Figs.~\ref{Fig2}(b,d).
Indeed, using Eqs.~\eqref{fre} with $\Delta=0$, one finds that 
this symmetric cycle is only possible if the period of 
the oscillation satisfies:
\begin{equation}
T_m= \frac{2D}{m}, \qquad \mbox{with ~ $m=1,3,\ldots$} 
\label{T}
\end{equation}
As parameters are varied, the reflection symmetry of the limit cycle
breaks down at a period-doubling bifurcation. 
Moreover, the inset of Fig.~3 shows that 
this bifurcation is followed by a 
period-doubling cascade as parameter $J$ is varied, 
giving rise to a state of collective chaos as that of 
Fig.~\ref{Fig2}(f).
Remarkably, though the collective dynamics is chaotic, 
the single-neuron evolution is not. 
Indeed, as a consequence of the mean-field character of the model and its first order kinetics,  
the firing order of the neurons 
is preserved (i.e.~neuron $j$ always fires just before neuron $j-1$) 
and mixing is not possible. 

\begin{figure}
\centerline{\includegraphics[width=70mm,clip=true]{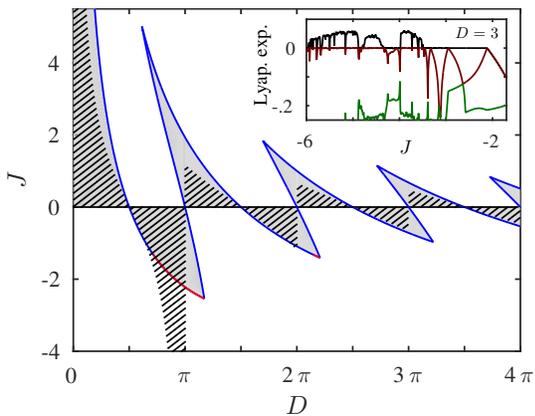}}
\caption{Stability regions of the incoherent and 
fully synchronized states for $\Delta=0$.
Shaded region: Incoherence is stable. Hatched region: Full synchrony is {\em unstable}.
Dark gray (blue) and the light gray (red) lines are the loci of sub- and super-critical
Hopf bifurcations of incoherence, respectively.
The approximate periodicity of the phase diagram with $D$ 
stems from the $\mathrm{ISI}=\pi$ of the uncoupled neurons.
Inset: Three largest (collective) Lyapunov exponents in the range $-6<J<-1.7$ for $D=3$, computed numerically from Eq.~\eqref{fre} using the method in \cite{farmer82}.
Note the supercritical Hopf bifurcation at $J_H^{(1)}=-2.116\ldots$ and the accumulation of period-doubling bifurcations at $J\approx3.5$.}
\label{Fig3}
\end{figure}

In the following, we analyze the FREs \eqref{fre} in detail, 
what permits to elucidate
why collective chaos is found only in a certain range of delays, 
and only for inhibitory coupling.  
For identical neurons, $\Delta=0$, the only fixed point is
$(r_s,v_s)=\left((J+\sqrt{J^2+4\pi^2})/(2\pi^2) ,0\right)$,
corresponding to an incoherent state.
Its stability can be determined linearizing around the fixed point
$r(t)=r_s+\delta r \,e^{\lambda t}$ and $v(t)=\delta v \, e^{\lambda t}$, 
and imposing the condition of marginal stability: $\lambda=i\Omega$.
We find a family of Hopf instabilities at
\begin{equation}
J_H^{(n)}=
\pi (\Omega_n^2-4) \times
\begin{cases}
(6\Omega_n^2+12)^{-1/2} \qquad \mbox{for odd $n$}
\\
(2\Omega_n^2-4)^{-1/2}  \qquad \mbox{for even $n$}   
\end{cases} 
\label{hopf}
\end{equation}
with associated frequencies $\Omega_n= n \pi/D$. 
The line with several cusps depicted in Fig.~\ref{Fig3} correspond to  
the boundaries of incoherence given by Eq.~\eqref{hopf}.  
The blue and red colors indicate the 
sub- and super-critical character of the bifurcation, respectively, 
and have been calculated perturbatively~\footnote{See Supplementary Material [url].}.
The stability region of incoherence (shaded) closely resembles 
that of the Kuramoto model of coupled oscillators, 
with alternating domains at positive and negative $J$ values 
as time delay is increased~\cite{YS99,CKK+00,MPS06,LOA09}. 
However, the presence of supercritical Hopf 
bifurcations in some ranges of the inhibitory part of the diagram 
is a distinct and important feature of model~\eqref{qif}-\eqref{etaj}, 
as we show below.

%%%%%%%%%%%%%%%%%%
%%%%%%%%%%%%%%%%% 

We also calculated the stability boundaries of the fully synchronized states, 
$V_j(t)=v(t)$,
which are given by the family of functions
\begin{equation}
J_c^{(m)}= 2 \cot\left(\frac{D}{m}\right), \qquad \mbox{with $m=1,3,5,\ldots$}
\label{fs}
\end{equation}
and by evenly spaced vertical lines at $D=n\pi$, with $n=1,2\dots$ 
~\footnote{See Supplementary Material [url].}.
Accordingly, the regions of \textit{unstable} full synchrony correspond to 
the hatched regions of the phase diagram Fig.~\ref{Fig3}.
Note that for weak coupling, i.e.~close to the $J=0$ axis, 
the phase diagram in Fig.~\ref{Fig3} is fully consistent with that of the Kuramoto model with delay~\cite{YS99},
as it can be proven applying the averaging approximation to model 
\eqref{qif} with $\Delta=0$, see~\cite{Kur84,PM14}.
Specifically, we observe three qualitatively different regions at small $|J|$: 
Incoherence (shaded-hatched), 
one or more fully synchronized states (white-unhatched), 
and coexistence between incoherence and full synchrony (shaded-unhatched).  

Away from the weak coupling regime, the system  
displays collective phenomena unseen in the Kuramoto model. 
Inside the unshaded-hatched region,
located below the Hopf curve $J_H^{(1)}$, both incoherence and synchronization 
are simultaneously \textit{unstable}.
Moreover, due to the supercritical character of 
the Hopf boundary $J_H^{(1)}$ in the range $2.250<D<3.684$ 
\footnote{Supercriticality is also present for the line $J_H^{(3)}$ in a tiny interval 
above $D=2\pi$},    
QPS emerges as a stable, small-amplitude oscillatory solution ---as that of 
Fig.~\ref{Fig2}(a)--- bifurcating
from incoherence with period $T=2D$. 

\begin{figure}
\centerline{\includegraphics[width=70mm,clip=true]{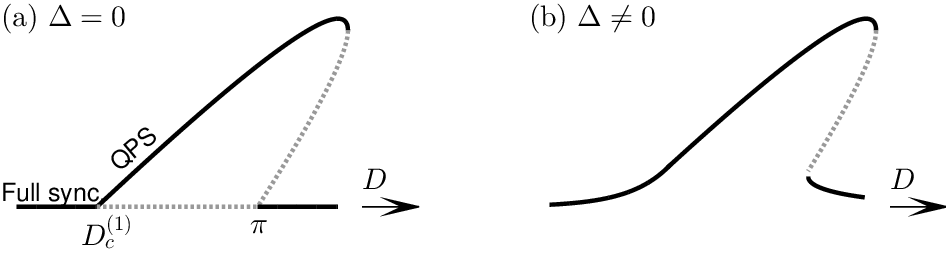}}
\caption{(a) Sketch of the transition between full synchronization and
QPS as $D$ is changed ($J<-2.54$, $\Delta=0$);
the $y$-axis is an arbitrary coordinate like e.g.~the  
 minimal value attained by the firing rate.
Full synchrony undergoes two transcritical bifurcations. The first one, at
$D_c^{(1)}$, gives rise to QPS which may, eventually, undergo secondary instabilities
(not depicted) leading to chaos. At $D=\pi$ full
synchrony recovers its stability abruptly. (b) Same sketch for $\Delta\ne0$.
Left transcritical bifurcation evaporates while
the right one becomes a saddle-node bifurcation. 
}
\label{Fig4}
\end{figure}

Additionally, QPS can also emerge via the destabilization of full 
synchronization at $J_c^{(1)}$. 
The simulation of the FREs confirms the prediction 
of Eq.~\eqref{fs},
and allows to complete a somewhat peculiar picture:
The fully synchronous state is a degenerate, infinitely long trajectory
along the $v$-axis, and the limit cycle corresponding to QPS
emanates from it with an unbounded size ---see Fig.~\ref{Fig2}(d), 
for a situation not far away from the bifurcation point.
In Fig.~\ref{Fig4}(a), a sketch of the bifurcation diagram 
(valid for $J<-2.54$ and $D$ around $\pi$) is depicted.
Stable QPS bifurcates from the fully synchronous state
at $D_c^{(1)}=\arctan(2/J) < \pi$, through a 
transcritical bifurcation of limit cycles.
Then, in a second transcritical bifurcation at $D=\pi$
(involving unstable QPS), the synchronized state recovers 
its stability. This scenario implies the existence of a 
region of bistability between QPS (or collective chaos) and full 
synchronization for $D>\pi$ ---in consistence, again, with the 
supercritical character of the Hopf bifurcation $J_H^{(1)}$ for $D<3.684$.

%%%%%%%%%%%%%%%%%%
%%%%%%%%%%%%%%%%%

So far, we have concentrated on identical QIF neurons. Our final results 
concern the robustness of QPS and collective chaos against  
heterogeneity. In the presence of heterogeneity full synchronization
and QPS cannot be observed, but states reminiscent of them persist, as sketched in Fig.~\ref{Fig4}(b).
Indeed, as the transcritical bifurcation is fragile, the bifurcation 
originally located at 
$D=\pi$ is replaced by a saddle-node bifurcation, whereas the  
other bifurcation at $D=D_c^{(1)}$ vanishes.

\begin{figure}
\centerline{\includegraphics[width=70mm,clip=true]{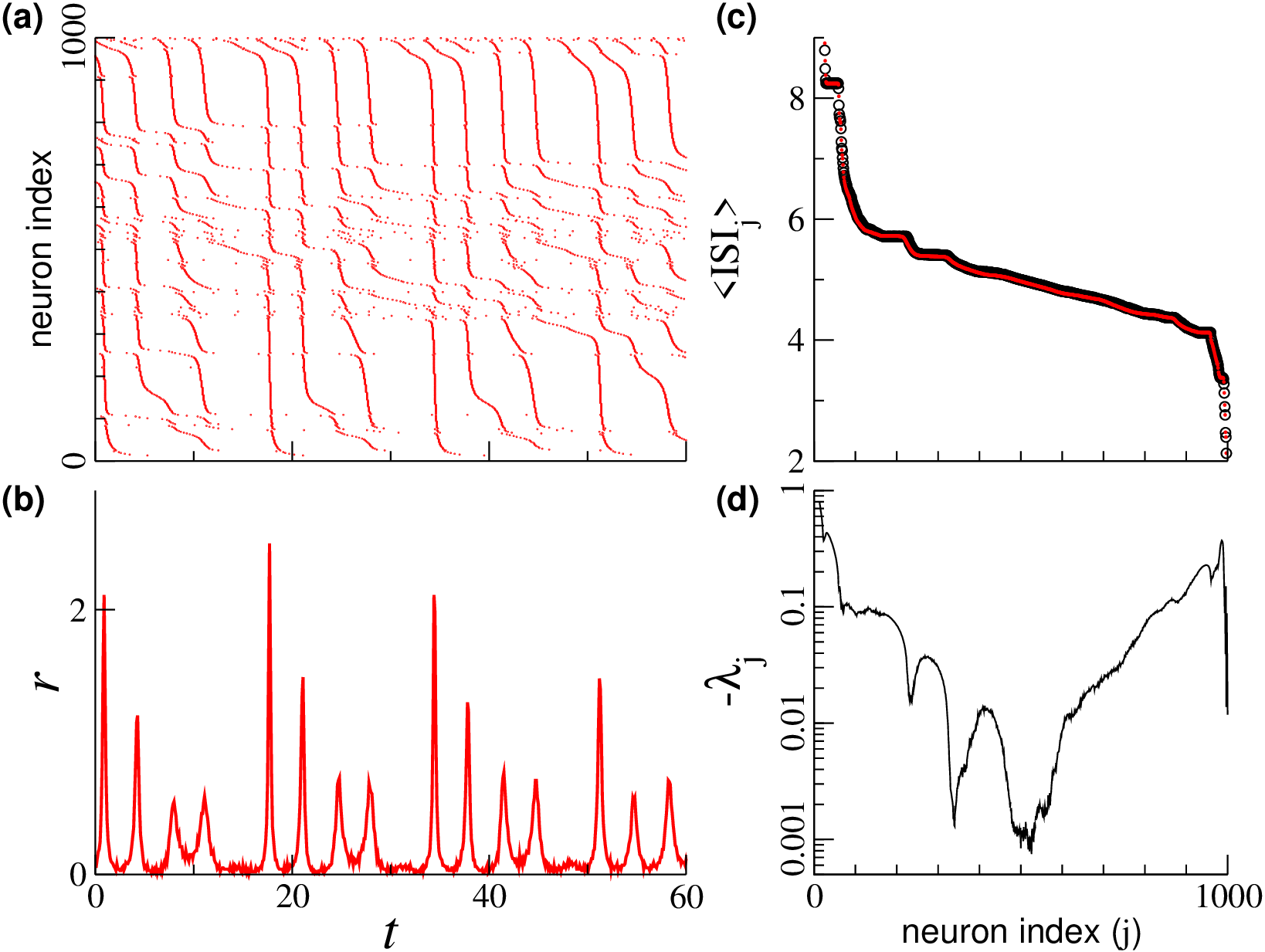}}
\caption{(a) Raster plot and (b) Time series of 
the mean firing rate of a population
of $N=1000$ heterogeneous ($\Delta=0.025$) QIF neurons, with
inhibitory coupling ($J=-3.8$) and delay ($D=3.5$). (c) Time-averaged 
ISIs of individual neurons vs.~neuron index (red points)---labels are  
associated to values of $\eta_j$. As a double-check,
the FREs~\eqref{fre} are simulated for the same parameters and used to drive individual neurons.
We obtain in this way the $\langle {\mathrm{ISI}}_j \rangle$
depicted by black circles. 
Note the horizontal segments corresponding
to clusters of neurons with identical $\langle \mathrm{ISI} \rangle$.
(d) Lyapunov exponents $\lambda_j$ of individual neurons driven by 
FREs ~\eqref{fre}. The $\lambda_j$'s are all negative, while $\lambda_j=0$ for the $\Delta=0$ case in Fig.~1(e).}
\label{Fig5}
\end{figure}

Regarding collective chaos, Fig.~\ref{Fig5}(a-c) shows  
numerical simulations of the heterogeneous 
QIF neurons \eqref{qif}, with parameter values close to those of Fig.~\ref{Fig1}(e).
We observe in Fig.~\ref{Fig5}(c) synchronized clusters at different average ISIs.
Using the FREs \eqref{fre} we checked that 
(i) the macroscopic infinite-$N$ 
dynamics of the model is indeed chaotic with leading Lyapunov exponents $\{0.013,0,-0.036\}$;
(ii) the microscopic dynamics is stable as revealed by the Lyapunov exponents obtained
forcing each neuron by Eq.~\eqref{fre}, see Fig.~\ref{Fig5}(d).
Interestingly, a similar state  
was numerically uncovered in~\cite{LP10}, and its chaotic nature
was attributed to the presence of quenched heterogeneity. 
However our conclusion is quite the opposite:
the chaotic state in Fig.~\ref{Fig5} 
can be regarded as a perturbed version of the collective chaos in Figs.~1(e) and 2(e),
and therefore heterogeneity is not essential for observing collective chaos.

%%%%%%%%%%%%%%%%%%
%%%%%%%%%%%%%%%%%

The fact that the model studied here exhibits nontrivial dynamics precisely
for inhibitory coupling ---in contrast to the previous studies using 
LIF neurons~\cite{Vre96,MP06,OPT10,LOP+12,BSV+14}---, deserves to be emphasized. 
A large body of data demonstrate that brain oscillations in the gamma 
and fast frequency ranges (30-200~Hz) 
are inextricably linked to the behavior of populations of 
inhibitory neurons~\cite{WB96,Wan10,WTK+00,BVJ07}. 
Moreover, theoretical and computational studies indicate that these oscillations 
emerge as a consequence of the interplay between inhibition and  
the significant time delays produced by synaptic processing, 
see e.g.~\cite{WB96,BH08}.
Our results add to these body of work, showing that 
QPS and collective chaos also arise in simple inhibitory 
populations of phase oscillators with 
delayed pulse coupling. Plausible values for the synaptic delays
are of the order of $D \sim 5$~ms, so that the QPS 
state studied here necessarily has a frequency $f \sim (2D)^{-1}=100$ Hz, 
corresponding to fast brain oscillations. 
This is in agreement with the frequency of the oscillations 
displayed by \textit{heuristic} firing rate models 
with fixed time delays and inhibitory coupling
~\cite{BH08,RBH05,BBH07,RM11}. Exactly 
the same range of frequencies is also observed in networks of identical, 
noise-driven inhibitory neurons with synaptic delays,
in the so-called sparsely synchronized state~\cite{BH08,BH99,BW03,BH06}. 
Remarkably, sparse synchronization also displays a 
macroscopic/microscopic dichotomy, 
similar to that of the QPS and collective-chaos states analyzed here.

The analysis of the thermodynamic limit of the model \eqref{qif}-\eqref{etaj} 
by means of the firing-rate equations \eqref{fre},
permits to dissect macroscopic from microscopic dynamics in that limit.
This strategy seems to be particularly useful 
for investigating collective chaos 
\cite{HR92,NK93,NK94,NK95,TGC09,TCG+11,KGO15}
as well as irregular activity states in heterogeneous neuronal
ensembles~\cite{LP10,UP16}.

\begin{acknowledgments}
We thank Hugues Chat\'e for valuable comments.
DP acknowledges support by MINECO (Spain) under the Ram\'on y Cajal programme.
We acknowledge support by MINECO (Spain) under project No.~FIS2014-59462-P, 
and by the European Union's Horizon 2020 research and innovation
programme under the Marie Sk{\l}odowska-Curie grant agreement No.~642563.
\end{acknowledgments}

% \bibliography{bibliografia}


\end{document}